\documentclass[a4paper,fleqn,usenatbib,letters]{mnras}

\usepackage{newtxtext,newtxmath}

\usepackage[T1]{fontenc}
\usepackage{ae,aecompl}

\usepackage{graphicx}	
\usepackage{amsmath}	
\usepackage{amssymb}	
\newcommand\SAL{Soviet Astronomy Letters}

\title[Cusp-filling tori oscillations]{MHD simulations of oscillating cusp-filling tori around neutron stars -- missing upper kHz QPO}

\author[V. Parthasarathy et al.]{Varadarajan Parthasarathy$^{1}$\thanks{E-mail: varada@camk.edu.pl}, W{\l}odzimierz Klu{\'z}niak$^{1,2}$\thanks{E:mail: wlodek@camk.edu.pl}, Miljenko \v{C}emelji{\'c}$^{1}$\thanks{E-mail: miki@camk.edu.pl}, 
\\
$^{1}$Copernicus Astronomical Center, ul. Bartycka 18, 00-716 Warszawa, Poland \\
$^{2}$KITP, University of California Santa Barbara, CA 93106, USA
}

\date{Accepted XXX. Received YYY; in original form ZZZ}

\pubyear{2017}

\begin{document}
\label{firstpage}
\pagerange{\pageref{firstpage}--\pageref{lastpage}}
\maketitle

\begin{abstract}
We performed axisymmetric, grid-based, ideal magnetohydrodynamic (MHD) simulations of oscillating cusp-filling tori orbiting a non-rotating neutron star. A pseudo-Newtonian potential was used to construct the constant angular momentum tori in equilibrium. The inner edge of the torus is terminated by a \textquotedblleft{}cusp\textquotedblright{} in the effective potential. The initial motion of the model tori was perturbed with uniform sub-sonic vertical and diagonal velocity fields. As the configuration evolved in time, we measured the mass accretion rate on the neutron star surface and obtained the power spectrum. The prominent mode of oscillation in the cusp torus is the radial epicyclic mode. It would appear that vertical oscillations are suppressed by the presence of the cusp. From our analysis it follows that the mass accretion rate carries a modulation imprint of the oscillating torus, and hence so does the boundary layer luminosity.
\end{abstract}




\section{Introduction}
\label{sec:one}

Accreting neutron stars (NS) in low mass X-ray binaries (LMXBs) manifest quasi-periodic oscillations (QPOs), pairs of which appear at kHz frequencies \citep{1996ApJ...469L...1V,1996ApJ...469L...9S,2000ARA&A..38..717V}. There are different regions that produce NS X-ray emissions: the optically thick disk, the hot corona above the disk, and the equatorial belt of accretion on the stellar surface, the boundary layer (BL), if the NS magnetic dipole is not sufficiently strong to channel accretion onto poles. Phase resolved spectroscopy reveals that the X-ray luminosity in QPOs originate from the boundary layer on the neutron star surface \citep{2003A&A...410..217G,2005AN....326..812G}. On theoretical grounds, low-frequency QPOs have also been attributed to the BL \citep{1999ApJ...518L..95T}, but in this contribution we only discuss the twin kHz QPOs. 

Several neutron star models have radii smaller than the innermost stable circular orbit (ISCO), such that the disk is decoupled from the neutron star surface \citep{1985ApJ...297..548K}. The accreted fluid releases a significant amount of energy in the boundary layer. It was shown that between 69$\%$ and 86$\%$ of the total luminosity is emitted from the boundary layer when matter falls from the ISCO onto the neutron star \citep{1986SvAL...12..117S}. The variable flow rate of the matter leaving the disk and crossing the \textquotedblleft{}relativistic accretion gap\textquotedblright{} between the ISCO and the neutron star \citep{1985ApJ...297..548K} will lead to an intensified luminosity modulation as the matter enters the boundary layer \citep{1987Natur.327..303P}. The frequencies of inner disk oscillations should thus lead to the variability of the boundary layer X-ray flux \citep{2005tsra.conf..604K,2007AcA....57....1A}.

The oscillations of neutron star accretion disks may modulate the X-ray flux by a general relativistic effect, the overflow of \textquotedblleft{}Roche lobe\textquotedblright{} (or a cusp-filling torus) near the ISCO. The overflow is achieved by terminating the inner edge of the torus by a 
\textquotedblleft{}cusp\textquotedblright{}, which is the point of self-crossing equipotential surface for any stationary rotating fluid with given angular momentum ($\ell_{\rm{c}}$). A cusp-filling torus accretes onto the star. The accretion is driven by gravity and pressure gradients, without need of viscous phenomena to transport the angular momentum \citep{1977ApJ...216..822P,1978A&A....63..221A}. An analytic model of mass accretion rate modulation as the fluid flows through the cusp of a torus oscillating in two axisymmetric eigenmodes has been constructed in a pseudo-Newtonian formulation \citep{2007AcA....57....1A}, and elaborated in general relativity for realistic equations of state to explain the twin peak QPOs observed in low mass X-ray neutron star binaries \citep{2016MNRAS.457L..19T}.

The recently launched astronomy satellite ASTROSAT, equipped with the Large X-ray Proportional Counters (LAXPC), will provide better opportunities to investigate the QPOs in LMXBs with accreting neutron stars. The LAXPC has the largest effective area in the hard X-ray band and is capable of high time resolution X-ray measurements in the 2$-$80 keV band \citep{2013IJMPD..2241009P}. 

We have used the PLUTO code \footnote{Freely available at http://plutocode.ph.unito.it/} \citep{2007ApJS..170..228M} to perform axisymmetric, grid-based ideal magnetohydrodynamic (MHD) simulations of oscillating cusp-filling tori around neutron stars. In order to mimic the effects of general relativity we used a pseudo-potential obtained by \citet[hereafter KL]{2002MNRAS.335L..29K}, which reproduces the ratio of orbital and epicyclic frequencies for the Schwarzschild metric. KL potential has been used by \citet{2016MNRAS.458..666P} in the recent numerical investigation of QPOs in perturbed tori orbiting a non-rotating black hole. 

The Letter is structured as follows. In Sect.~\ref{sec:two} we explain the numerical set-up and present the results of our simulations. In Sect.~\ref{sec:three} we discuss the results and summarize in Sect.~\ref{sec:four}.

\section{Simulations and Results}
\label{sec:two}
\begin{figure*}
\centering
\includegraphics[height=0.65\columnwidth, width=1.0\columnwidth]{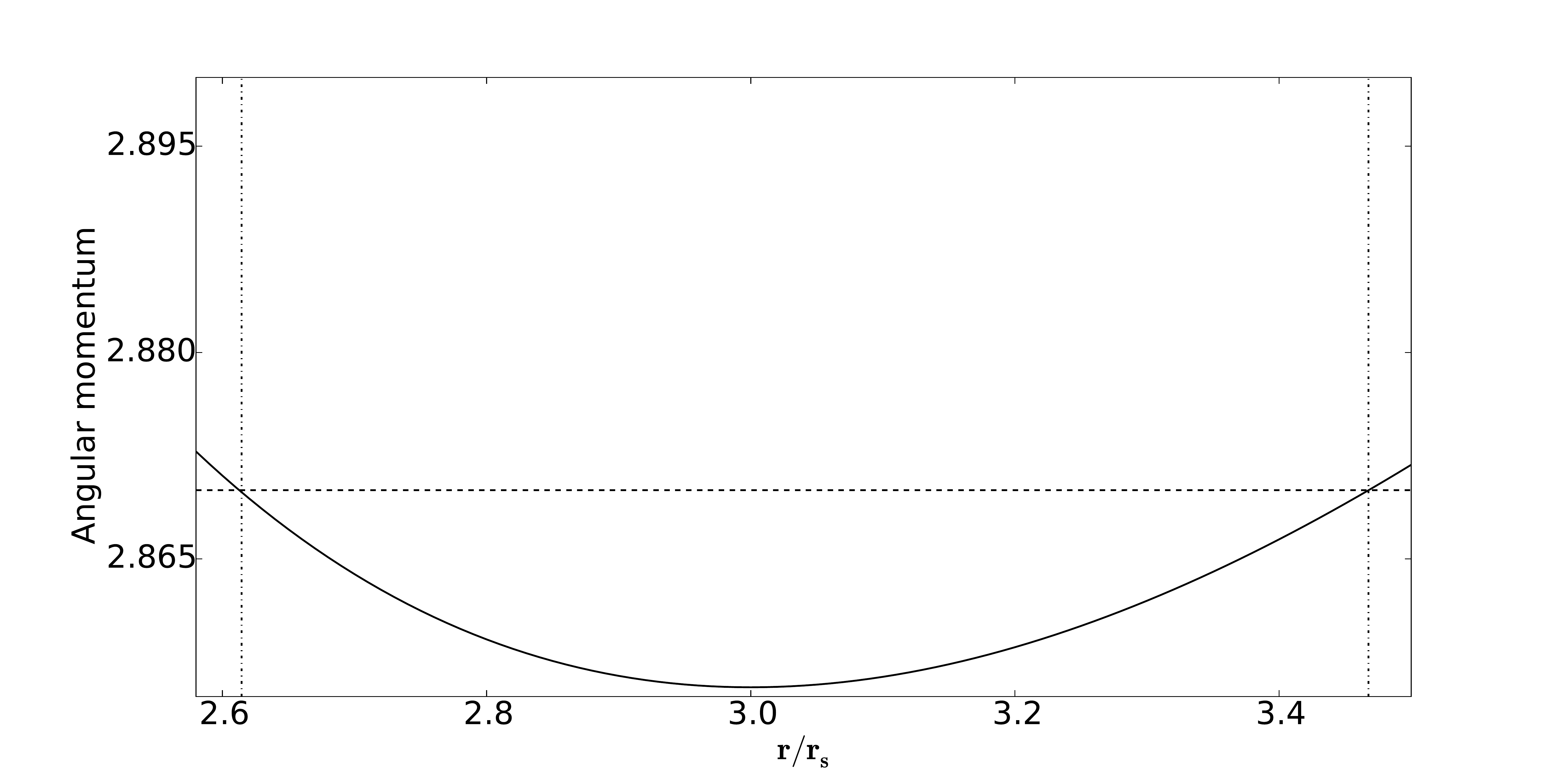}
\includegraphics[height=0.6\columnwidth, width=0.9\columnwidth]{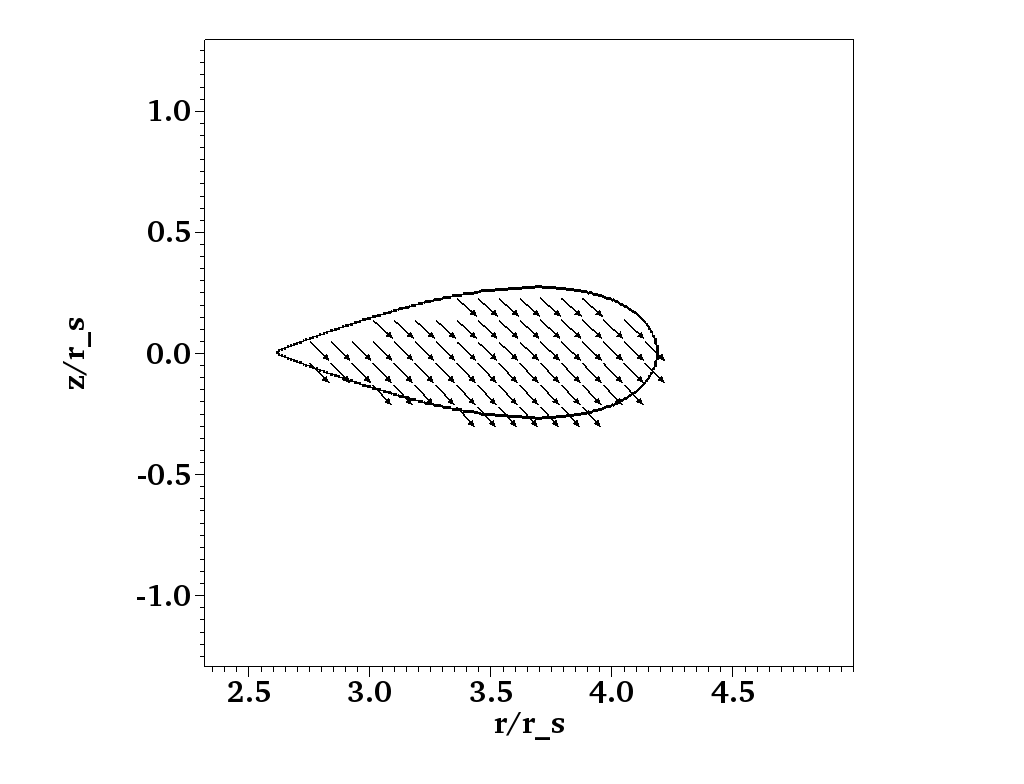}
\caption{Left panel: The Keplerian angular momentum (solid line) and the angular momentum in the flow (thick dotted line). The location of cusp ($r_{\rm{cusp}}$) is at 2.62 $r_{\rm{s}}$ (thin dotted line) and the center of torus ($r_{\rm{c}}$) is at 3.47 $r_{\rm{s}}$ (thin dotted line). Right panel: Initial configuration of the cusp-filling torus for model IMHDD1. Solid line represent the density contour in a meridional cross-section of the cusp-filling torus at $r_{\rm c}$ = 3.47 $r_{\rm{s}}$. The inner edge of the torus is terminated by a cusp at 2.62 $r_{\rm{s}}$. Solid arrows represents the initial uniform sub-sonic diagonal velocity perturbation.}
\label{fig:ini}
\end{figure*}

We consider a torus accreting onto a canonical neutron star of mass 1.4 $M_{\odot}$ and radius 10 km. The numerical set-up and velocity perturbations in spherical coordinates ($r, \theta$) are adapted from \citet{2016MNRAS.458..666P}. The surface of the neutron star is at 2.4 $r_{\rm{s}}$, where the Schwarzschild radius is $r_{\rm{s}} = $ 2 $GM/c^{2}$. The outer radial domain is at 5 $r_{\rm{s}}$. The cusp of the torus is located at $r_{\rm{cusp}} = 2.62 r_{\rm{s}}$ and the radius of pressure maximum ($r_{\rm{c}}$) is at 3.47 $r_{\rm{s}}$ (see Fig.~\ref{fig:ini}). The inner and outer co-latitudinal boundaries are located at  $(\pi/2 - 0.3)$  and $(\pi/2 + 0.3)$, respectively.  A description of the models is given in Tab.~\ref{table:model}.

A radially increasing logarithmic grid is used, which preserves the cell aspect ratio. In such a grid the intrinsic characteristic spatial scale grows linearly with the distance from the stellar surface, exactly as the grid spacing of latitude. The grid resolution introduces about the same amount of numerical dissipation at all radii. Uniform grid is used in the latitudinal direction. The inner radial boundary is prescribed with a pseudo-reflective condition for modelling the neutron star surface, in which the components of the velocity and magnetic fields normal to the cell interface do not flip the signs. 

\begin{table}
\caption{A list of models with model name, resolution, strength of the magnetic field and  the initial profile of the magnetic field. The type of velocity perturbation is also listed.}
\label{table:model}
\begin{tabular}{c|c|c|c|c}
\hline
Model & Resolution  & $B_{\rm{NS}}$ & $B_{\rm{NS}}$ profile & Perturbation \\
\hline
IHDV    & 420 $\times$ 300  & $-$ & $-$ & vertical \\
IHDD    & 420 $\times$ 300  & $-$ & $-$ & diagonal \\
IMHDD1  & 420 $\times$ 300  & 900 $\rm{G}$  & dipole & diagonal \\
IMHDD2  & 140 $\times$ 100  & 9000 $\rm{G}$ & dipole & diagonal \\
\hline
\end{tabular}
\end{table}

In order to imitate the effects of general relativity we use the KL potential, 
\begin{equation}
\Phi_{\rm KL} = -\frac{GM}{3r_{\rm s}}\left(e^{3r_{\rm s}/r} - 1\right),
\label{eqn:kl}
\end{equation}
where $r$ is the distance from the center of the neutron star. The KL potential reproduces the ratio of orbital and epicyclic frequencies for the Schwarzschild metric
\begin{equation}
    \omega_{\rm{R}} = \left(1 - \frac{3r_{\rm{s}}}{r}\right)^{1/2} \Omega_{\rm{K}},
    \label{eqn:rad}
\end{equation}
where $\omega_{\rm{R}}$ is the radial epicyclic frequency and $\Omega_{\rm{K}}$ = 2$\pi \nu_{\rm{K}}$ is the orbital frequency (see also \citet{giu,vw}). The eigenfrequency of the radial epicyclic mode as a function of the equilibrium radial position of the torus ($r_{\rm{c}}$), is equal to 0.36 in units of Keplerian frequency ($\nu_{\rm{K}}$ at $r_{\rm{c}}$). In a spherically symmetric potential, the vertical epicyclic frequency is equal to $\Omega_{\rm{K}}$. 

We set-up a stationary hydrodynamic cusp-filling torus around a non-rotating neutron star in the limit of vanishing magnetic dipole fields in the manner described hereafter. A constant angular momentum torus in  equilibrium is initialised around the central compact body prescribed with KL potential (see Eq.~\ref{eqn:kl}). The inner radius of the torus coincides with the point of self-crossing equipotential surface ($r_{\rm cusp}$). The initial motion of the cusp-filling torus is unperturbed. There is a steady outflow of matter through the cusp of the stationary torus onto the star. The evolved quasi-equilibrium state of the accreting cusp-filling torus is used as the initial configuration for the hydrodynamic models (see Tab.~\ref{table:model}). The motion of the evolved torus is triggered with uniform sub-sonic vertical and diagonal velocity fields to achieve the models IHDV and IHDD, respectively. The MHD models (see Tab.~\ref{table:model}) include a weak NS dipole, and are perturbed with uniform sub-sonic diagonal velocity fields from the beginning.

\begin{figure*}
\includegraphics[width=18cm,height=6cm]{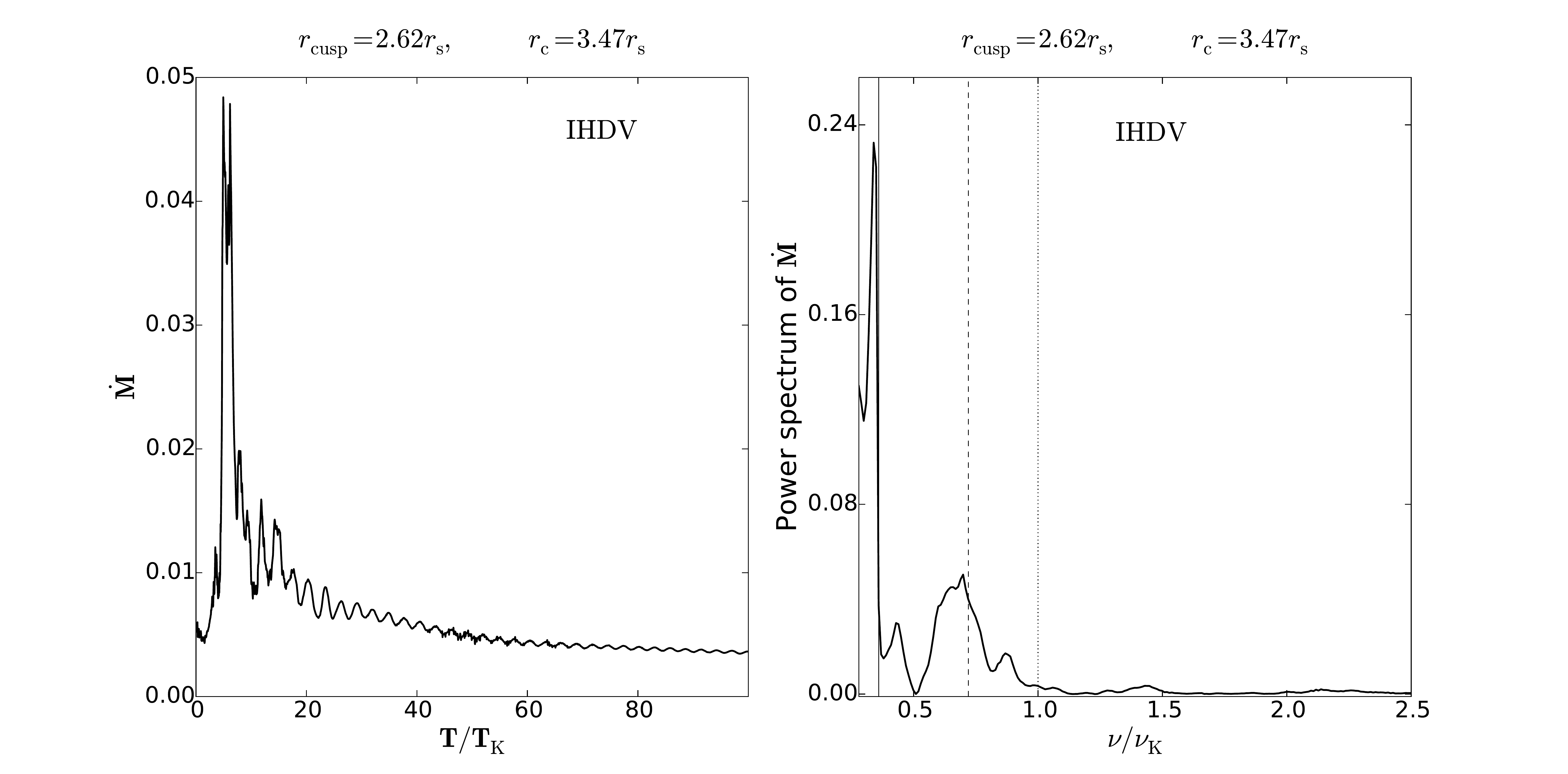}

\vspace{-0.1cm}

\includegraphics[width=18cm,height=6cm]{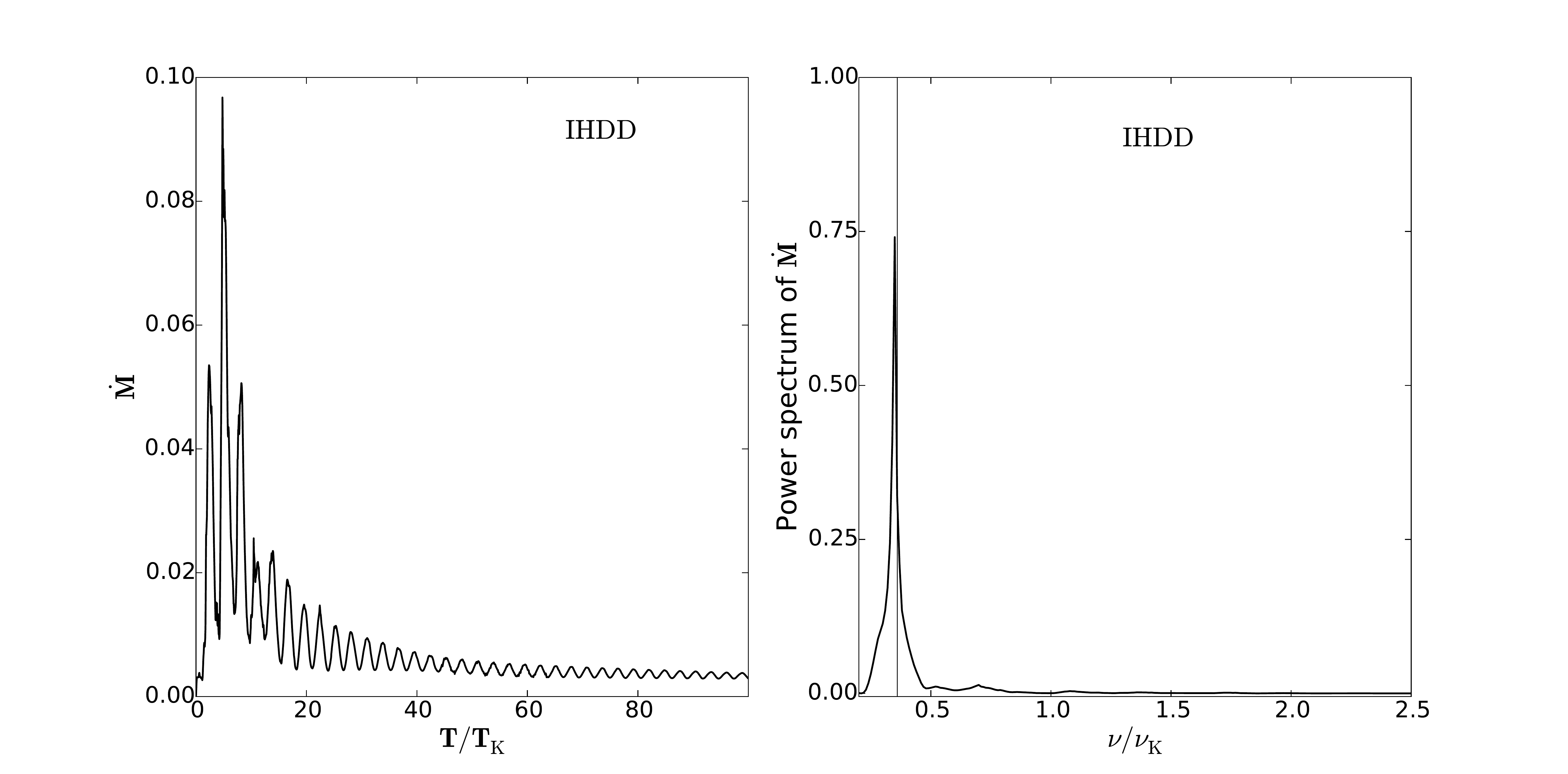}
\caption{Left panels: Mass accretion rate on the neutron star surface as a function of Keplerian time $T_{\rm{K}}$ at center of torus $r_{\rm{c}}$ for models IHDV and IHDD, in the units of $\dot{M} = 1.6 \times 10^{-10} M_{\odot} \rm{yr}^{-1}$. Right panels: Power spectrum of $\dot{M}$ measured on the neutron star surface for models IHDV and IHDD. Frequencies are in units of Keplerian frequency $\nu_{\rm{K}}$ at center of torus $r_{\rm{c}}$. The solid and dotted lines corresponds to the theoretical frequency values of the radial and vertical eigenmodes of slender tori at $r_{\rm{c}}$, respectively. The eigenfrequencies of the radial and vertical epicyclic modes of the oscillating tori are 0.36 $\nu_{\rm{K}}$ and 1.0 $ \nu_{\rm{K}}$ at $r_{\rm{c}}$, respectively. The dashed line correspond to the first harmonic of the radial mode of the torus. Additional frequencies may also be present in the simulations.}
\label{fig:hd}
\end{figure*}

\begin{figure*}
\includegraphics[width=18cm,height=6cm]{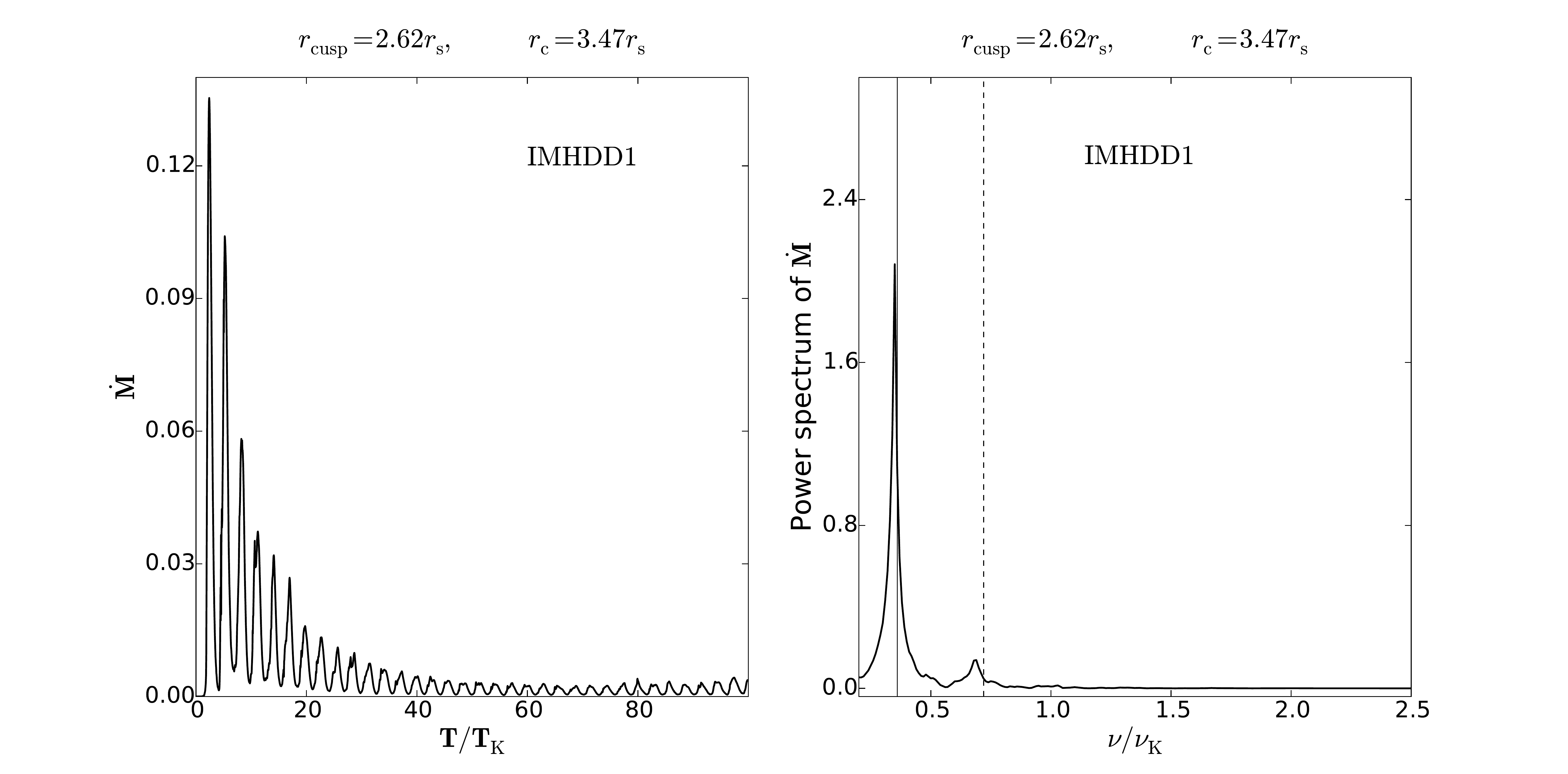}

\vspace{-0.1cm}

\includegraphics[width=18cm,height=6cm]{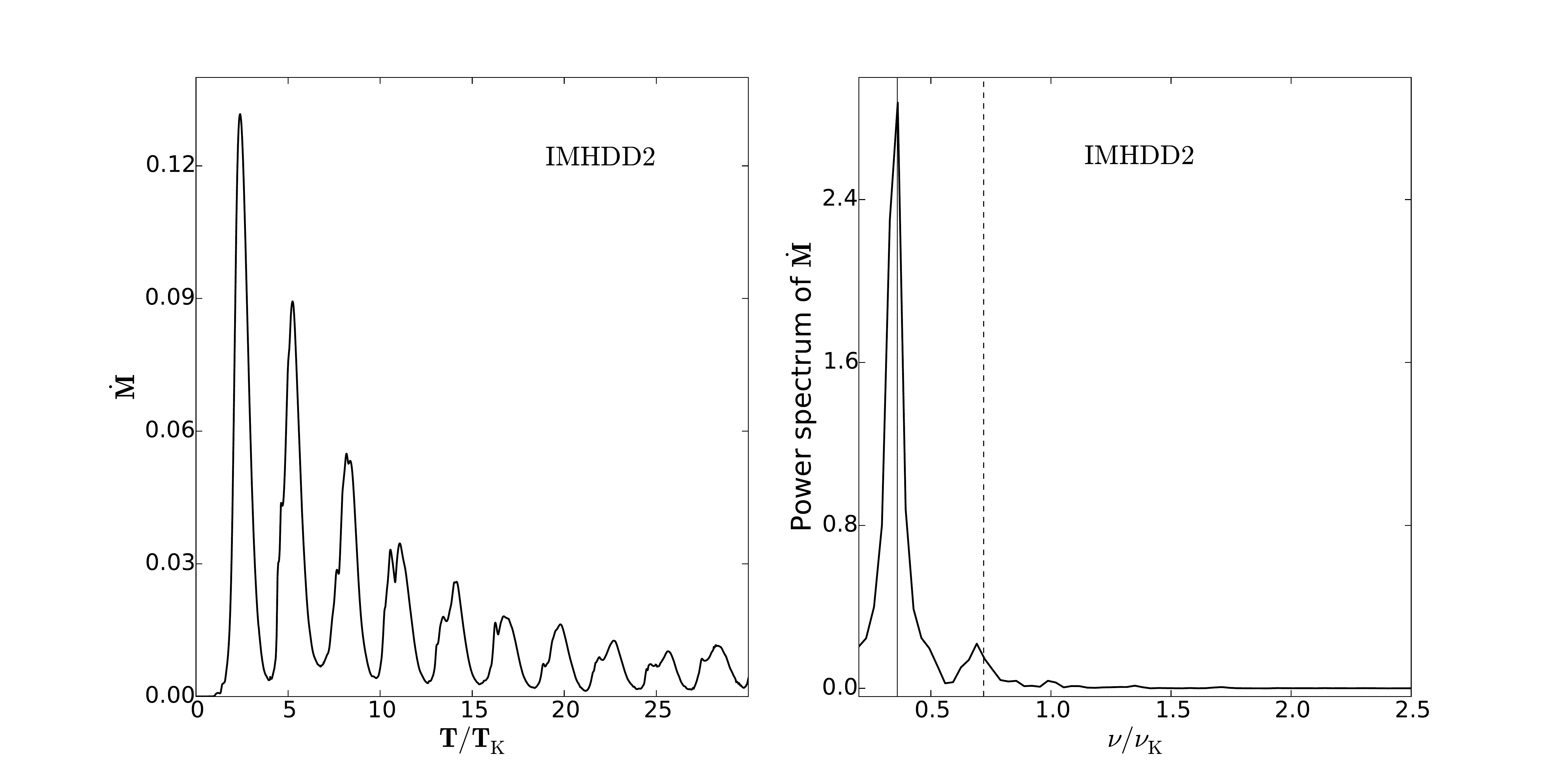}
\caption{Same as Fig.~\ref{fig:hd} for models IMHDD1 and IMHDD2.}
\label{fig:mhd}
\end{figure*}

We measured the mass accretion rate ($\dot{M}$) on the neutron star surface and obtained the power spectrum of $\dot{M}$ using the fast Fourier transform (FFT) technique. The power spectrum of $\dot{M}$ for the equilibrium model of the stationary overflow of Roche lobe does not manifest peaks that correspond to a periodicity. The results of models IHDV and IHDD are shown in Fig.~\ref{fig:hd}, and of models IMHDD1 and IMHDD2 in Fig.~\ref{fig:mhd}. The most prominent mode excited with the vertical and diagonal perturbations (see Tab.~\ref{table:model}) is the radial epicyclic mode of the cusp torus. The first harmonic of the radial mode is clearly visible in the power spectrum of the models IHDV (Fig.~\ref{fig:hd}), IMHDD1 and IMHDD2 (Fig.~\ref{fig:mhd}). Additional frequencies are seen in the simulation of the model IHDV (Fig.~\ref{fig:hd}).


\section{Discussion}
\label{sec:three}
We numerically investigate the modulation of the neutron star boundary layer luminosity by constructing a torus with a constant distribution of angular momentum in the KL potential (Fig.~\ref{fig:ini}). The fluid outflows from the torus through the cusp. The models of cusp-filling tori were initially perturbed with uniform sub-sonic vertical and diagonal velocity fields (Tab.~\ref{table:model}). The results of our simulations are discussed below.  

The model IHDV has been perturbed with uniform sub-sonic vertical velocity perturbation. The restoring force of a vertical oscillation is the vertical gradient of the gravitational potential. Since the cusp corresponds to the self-crossing of the equipotential surface, the gradient of the effective potential vanishes at the cusp. This seems to be suppressing vertical motion in the inner parts of the torus. The prominent mode of oscillation excited by the vertical motion of the model IHDV is the radial epicyclic mode of the torus (Fig.~\ref{fig:hd}). In the model IHDV we observe the first harmonic of the radial mode and additional frequencies.  

\citet{2007AcA....57....1A} analytically showed that the mass accretion rate retains the modulation imprint of the vertically oscillating disk, at double the mode frequency. \citet{2016MNRAS.458..666P} numerically studied the frequencies of oscillation of a torus with an initial vertical motion, and demonstrated that the power of vertical motion in the density $L_{2}$ norm is concentrated in the first harmonic at twice the orbital frequency. In both the cases this is related to the fact that the equatorial plane is crossed twice per every vertical oscillation cycle. We expected to see a prominent vertical oscillation in the model IHDV. However, it was not observed in the $\dot{M}$ measured on the neutron star surface and the corresponding power spectrum of $\dot{M}$ (see model IHDV in Fig.~\ref{fig:hd}). The uniform initial velocity perturbation does not excite the vertical eigenmode of the cusp-filling torus, since there is no restoring force at the cusp.

The models IHDD, IMHDD1 and IMHDD2 have been perturbed with uniform sub-sonic diagonal velocity perturbations. The diagonal perturbation may excite the radial and vertical eigenmodes \citep{2016MNRAS.458..666P,2017MNRAS.467.4036M}. We infer the radial epicyclic mode of the torus from the diagonal perturbations of the models IHDD (Fig.~\ref{fig:hd}), IMHDD1 and IMHDD2 (Fig.~\ref{fig:mhd}), respectively. The vertical eigenmode was not excited in the simulations of models IHDD, IMHDD1 and IMHDD2.  The first harmonic of the radial mode is present in the power spectrum of the models IMHDD1 and IMHDD2 (Fig.~\ref{fig:mhd}). 

The results obtained from all the cusp-filling tori models are the same, although the initial conditions were not identical in the HD and MHD regimes (see Sec.~\ref{sec:two}). The most prominent peak in the power spectrum of vertically and diagonally perturbed tori (see Figs.~\ref{fig:hd} and ~\ref{fig:mhd}) is the radial epicyclic mode frequency. The conclusion of our numerical analysis is robust in showing the modulation of the boundary layer luminosity by disk oscillations and a strong presence of radial epicyclic frequency in all models. However, the vertical epicyclic frequency is suppressed.

The variation of the quality factor and amplitude of the upper and lower kHz QPOs in a low-mass X-ray binary source was studied in \citet{2005MNRAS.357.1288B,2005MNRAS.361..855B}. They analysed the archival data from the \textit{Rossi X-ray Timing Explorer} (RXTE). The quality factor gives a measure of the coherence of the oscillator. It was shown that the upper and lower QPOs follow different paths in the quality factor versus frequency diagram. The lower kHz QPOs exhibited much higher coherence (higher Q-factor) than the upper kHz QPOs. If we assume that in low-mass X-ray binary sources, the vertical eigenmode correspond to the upper kHz QPOs and the lower kHz QPOs to the radial epicyclic mode, our results seem to be in qualitative agreement with this finding. The upper kHz QPO is hard to discern in the boundary layer luminosity of our simulation. In contrast, the radial epicyclic frequency is a robust feature in the power spectrum of the mass outflow from the cusp-filling torus, appearing even when the initial perturbation is purely vertical.

\section{Conclusions}
\label{sec:four}
Using the PLUTO code we performed axisymmetric ideal MHD simulations of oscillating cusp-filling tori around non-rotating neutron stars in the KL potential. For the first time we numerically study the modulation of the mass accretion rate on the neutron star surface. We draw the inference that, both the vertical and diagonal excitations of cusp-filling tori excite the radial epicyclic mode in such tori. Based on \citet{2016MNRAS.458..666P}, we expected that the vertical epicyclic mode may be excited, however the uniform velocity perturbations fail to excite the vertical eigenmode of the cusp-filling torus. Our simulations confirm the modulation mechanism of the neutron star boundary layer luminosity by disk oscillations \citep{2007AcA....57....1A}. However, the vertical eigenmode is suppressed as there is no restoring force at the cusp. The strong presence of the radial epicyclic frequency in the boundary layer luminosity (presumed to be the lower kHz QPO) and the absence of the vertical epicyclic frequency  (presumed to be the upper kHz QPO) may be interpreted as corresponding to the high Q-factor of the lower kHz QPO \citep{2005MNRAS.357.1288B,2005MNRAS.361..855B} and the low Q-factor of the upper kHz QPO \citep{2005MNRAS.361..855B}.

The results of our numerical analysis are relevant to the ASTROSAT mission equipped with LAXPC \citep{2013IJMPD..2241009P} which will investigate the quasi-periodic oscillations in low mass X-ray binaries with low magnetic field neutron stars. 

\section*{Acknowledgements}

The authors are grateful to the anonymous referee for the helpful comments. This research was supported by the Polish NCN grant 2013/08/A/ST9/00795, and in part by the NSF under grant No. NSF PHY11-25915.




\bibliographystyle{mnras}
\bibliography{modulation}


\bsp	
\label{lastpage}
\end{document}